\documentclass[twocolumn]{aastex62}

\usepackage{textcomp}
\usepackage{amsmath}

\submitjournal{Icarus (published in 339:113569 (2020))}

\begin{document}

\title{Lifting Grains by the Transient Low Pressure in a Martian Dust Devil}

\correspondingauthor{Tetyana Bila}
\email{tetyana.bila@stud.uni-due.de}

\author{Tetyana Bila}
\author{Gerhard Wurm}
\author{Florence Chioma Onyeagusi}
\author{Jens Teiser}

\affiliation{University of Duisburg-Essen}

\begin{abstract}
Lifting dust and sand into the thin Martian atmosphere is a challenging problem. Atmospheric pressure excursions within dust devils have been proposed to support lifting. We verify this idea in laboratory experiments. Pressure differences up to a few Pa are applied along particle layers of 50 to 400 $\rm \mu m$. As samples we used glass beads of $\sim$50 $\rm \mu m$ diameter and irregular basalt grains  of $\sim$20 $\rm \mu m$ in size. The total ambient pressure of air was set to 600 Pa. 
Particles are ejected at pressure differences as low as 2.0 $\rm \pm$ 0.8 Pa.
In the case of glass beads, the ejected grains returning to the particle bed can trigger new particle ejections as they reduce cohesion and release the tension from other grains. Therefore, few impacting grains might be sufficient to sustain dust lifting in a dust devil at even lower pressure differences. Particle lift requires a very thin, $\sim$100 $\rm \mu m$, low permeability particle layer on top of supporting ground with larger pore space.
Assuming this, our experiments support the idea that pressure excursions in Martian dust devils release grains from the ground.

\end{abstract}


\section{Introduction}

Dust devils and their tracks are frequently observed on Earth and Mars 
\citep{Fisher2005, Reiss2016, Lorenz2016, Hausmann2019}. 
But while on Earth gas drag of the moving vortex is sufficient to pick up sand and dust, curiously enough, Martian conditions are often at the edge of allowing particles to be lifted from the ground by gas drag alone. More generally, apart from dust devil research, there is a large amount of literature on entraining particles into the Martian atmosphere by gas drag, elaborating on observations, laboratory experiments, and simulations
\citep{Greeley1976, White1987, Greeley1980,Forget1999, Shao2000, Duran2011, Kok2012, Bridges2012, Rasmussen2015, Daerden2015, Zurek2017, Baker2018}


But in spite of all this work, lift by wind induced gas drag might still be debated. This is a motivation for our work but not the focus. Therefore, we do not dwell on its details and only note that wind will, in most cases, still provide one of the main lifting forces on dust grains at the Martian surface. Its disputed capabilities of lifting grains, however, inspired research on a number of supporting mechanisms.

An overview of proposed ways to support particle entrainement into dust devils is given by \citet{Neakrase2016}. These mechanisms range from charged grains in electrical fields \citep{Schmidt1998, Zheng2003, Merrison2004, Harrison2016, Franzese2018, Wurm2019}, over fluffy dust aggregates with much lower density and lower cohesive forces \citep{Merrison2007}, to thermoluminous effects \citep{Debeule2014,Debeule2015,Kuepper2016,Koester2017, Schmidt2017}, which are only existing at the low atmospheric pressure within insolated Martian soil and are not present on Earth.

Finally, also pressure drops coming along with a dust devil vortex have been proposed to drive particle entrainment \citep{Greeley2003,Balme2006}. It is this latter, which we will exploit further here.
This pressure drop mechanism is based on the following idea. The pressure within the porous soil is usually in equilibrium with the atmospheric pressure above. However, if a short pressure excursion occurs, the soil needs some time to adapt as the gas has to flow through soil, which, with a limited permeability, poses a large resistance to the flow. The gas needs a certain time before equilibrium is restored again. During the short transient time of a dust devil passage, there might be a pressure difference between some gas reservoir within the soil and the atmosphere above. As the vortex of a dust devil is a low pressure region, soil particles are subject to a lifting force as the dust devil moves along. This mechanism is often named $\Delta \rm P$-effect. 

\begin{figure}
\includegraphics[width=\columnwidth]{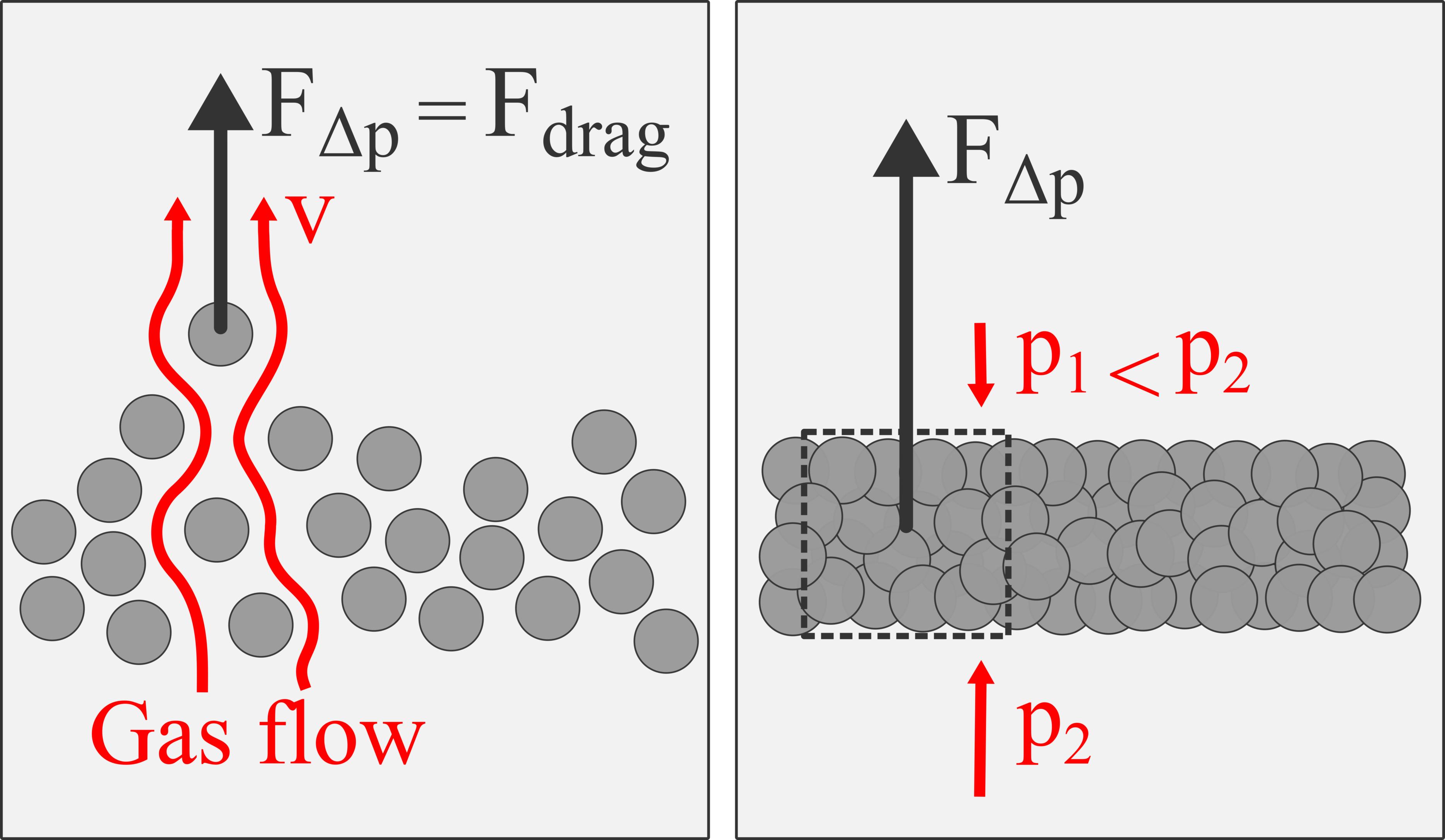}
    \caption{\label{fig.idee}$\Delta \rm P$-effect. Left: Gas drag due to a high
    efficiency flow at high permeability. This flow is driven by the pressure difference or $\Delta \rm P$. Right: Static pressure difference $\Delta \rm P$ provides lift directly}
\end{figure}

To estimate the potential of the $\Delta \rm P$-effect, \citet{Balme2006} considered two cases. If the pores of the top soil are large, there is a significant gas flow, i.e. gas with a significant speed driven by the pressure difference. In this case, the grains do not react to the pressure difference itself but the typical gas drag of a particle being immersed in a gas flow. The other extreme is the static force due to the pressure difference if the soil would be sealed perfectly gas tight. These ideas are sketched in fig. \ref{fig.idee}.

In a first attempt to test this model, \citet{Koester2017a} set up an experiment where they apply a pressure difference to different layers of dust and sand but with thickness on the cm-scale. They found lifting for these thick layers once the weight of the total layer was balanced by the pressure difference. In any case, the required absolute pressure differences for these thick layers were way too high on the kPa level to be directly applicable to Mars. 

In general, on Mars, the pressure drops only by a few Pa in a dust devil  \citep{Murphy2002,Ellehoj2010,Kahanpaeae2016,Steakley2016}. 
More precise statements might be given in the future as simultaneous observations and meteorological measurements of dust devils on Mars proceed and improve our understanding of low pressure vortices \citep{Reiss2018, Kahan2018}. 

\citet{Koester2017a} found that the pressure difference necessary to lift a layer of certain thickness is about 30 Pa/mm for the material they studied. Therefore, if pressure differences in a Martian dust devil should suffice, only grains in layers on the order of 100 $\rm \mu m$ can be lifted and only if the total pressure difference really drops only along this layer.

Such thin layers are not easily established though, neither in nature nor in the laboratory. They cannot exist as free, thin sheets of dust. This is virtually impossible as these sheets would readily collapse and find themselves on a supporting ground below. {So, a basic question is, what does this supporting ground look like?}
 
One stable option would be a thin dust layer on top of a sand layer. This would still count as a thin dust layer within the concept studied here, as the sand layer is much more permeable to the gas than the dust layer. Therefore, somewhat idealized, only the dust layer would be {resistant} to the gas flow and the pressure would drop only along this layer and not along the permeable, supporting material. On Mars this might be atmospheric dust of $\rm \sim 1$ $ \mu m$ in size \citep{Lemmon2015} that settled on top of sand sized soil. The top layer could then be of low porosity material with sand grains connected by dust grains and both types of grains might be lifted from a dust free sand ground below. 
 
To test this mechanism at this extreme small size scale of only 100 $ \rm \mu m$, assuming such layers and enough gas reservoir might exist, we created artificial thin layers of dust in the laboratory and tested, at what pressure differences along this layer particles are lifted. 

It might not be taken for granted that dust is lifted at all with low pressure differences, even if this is suggested by measurements on thick layers. 
It has to be taken into account that the thinner the layer gets, the more important cohesion might become. For thinner layers the surface to volume ratio for ejected layer parts becomes larger. The surface is responsible for cohesion and the volume for mass and weight. 

This is just a simplified view and
it is by far not clear how this works out in the total picture, as cohesion is rather variable. 
{According to \citet{Johnson1971} the cohesion between a spherical grain of radius $r$ and a flat wall is calculated as}

\begin{equation}
F_c = 3 \pi r \gamma,
\end{equation}

with the surface energy $\gamma$. 
On one side, the surface energy is more or less well known. Typical values found in the literature are e.g. on the order of $\gamma = 0.01$  $ \rm J/m^2$ or even larger for silicates \citep{Kimura2015, Steinpilz2019}. 
On the other side, wind tunnel experiments are often more consistent with a surface energy which is orders of magnitude lower or $\gamma = 10^{-4}$ $ \rm J/m^2$ \citep{Shao2000, Kruss2019, Demirci2019}. This is not a contradiction though. If e.g. very small grains are attached to larger ones or if particles are irregular and the curvature of a contact is much smaller than the overall grain size, the effective surface energy is reduced with respect to the large particle size. The effective surface energy is therefore highly variable and hard to predict. It makes all the difference though for lifting grains.

I.e. for comparing cohesion with weight, the gravitational force on a grain on Mars is 
\begin{equation}
F_g = \frac{4}{3} \pi r^3 \rho g_M. 
\end{equation}

with the particle density $\rho$ and the gravitational acceleration $g_M$.
Putting in numbers, e.g. a grain radius of $r = 100$ $ \rm  \mu m$,
a density of $\rho = 3000$ $ \rm kg/m^3$ and Martian gravity of $g_M = 3.7$ $ \rm m/s^2$ this is about $F_g = 5 \cdot 10^{-8}$ N.
For the surface energies given above, we either get $F_c = 10^{-5} $ N or $F_c = 10^{-7}$ N. So in one case, cohesion dominates over gravity by orders of magnitudes and grains can never be lifted just by a pressure difference. In the other case, cohesion is on the order of gravity. As scaled from \citet{Koester2017} gravity can be compensated by a few Pa pressure difference for a 100 $\rm \mu m$ layer. Therefore lift under Martian dust devil conditions might be possible in the case of low cohesion. From these estimates
one might expect some grains to be lifted from a thin dust layer at low pressure difference. However, it also has to be taken into account that the gas flow through only a few layers of grains is not homogeneous and might change the simple picture further.

In summary, it seems possible that grains are lifted but this is difficult to be estimated from calculations alone. 
We therefore did set up an experiment to verify if the $\Delta P$-effect is a viable mechanism at all. We did set up the best case Martian conditions as good as possible. We simulated conditions of a very small pressure drop along a very thin particle layer in a low ambient pressure.

\section{Experiment}

In principle, the experimental setup just follows the basic idea of the $\Delta P$-effect. An increasing pressure difference is applied to a particle layer of given thickness. The particles are observed and ejections are recorded with a camera and correlated to the pressure difference at the time of ejection. The preparation and observation of a rather thin particle layer and application of a really low pressure difference are the crucial part here. 

On Mars, grains being lifted settle again, eventually. We also prepared samples by having them sedimenting on top of a porous mesh. However, the particle bed cannot be prepared before the experiment has reached its final ambient pressure in contrast to \citet{Koester2017a}. Otherwise, as we aim for very low pressure differences and very thin layers of grains, the evacuation of the vacuum system generates too much pressure difference. Such samples get destroyed before a measurement can be taken. One could devise a procedure of very slow evacuation but we consider this to be not practical. We therefore only started to sieve grains onto the supporting mesh after the final ambient pressure was reached. The mesh has a slightly smaller or comparable spacing than the grains (basalt: 25 $\rm \mu m$, glass: 50 $\rm \mu m$) allowing larger grains to accumulate on top, but otherwise the space below is empty. While the size distribution of Martian dust and sand is certainly different, this procedure is otherwise similar, keeping in mind that the g-level on Earth is higher which can change the arrangement and effective cohesion of grains \citep{Musiolik2018}.

\begin{figure}
\begin{center}
\includegraphics[width=\columnwidth]{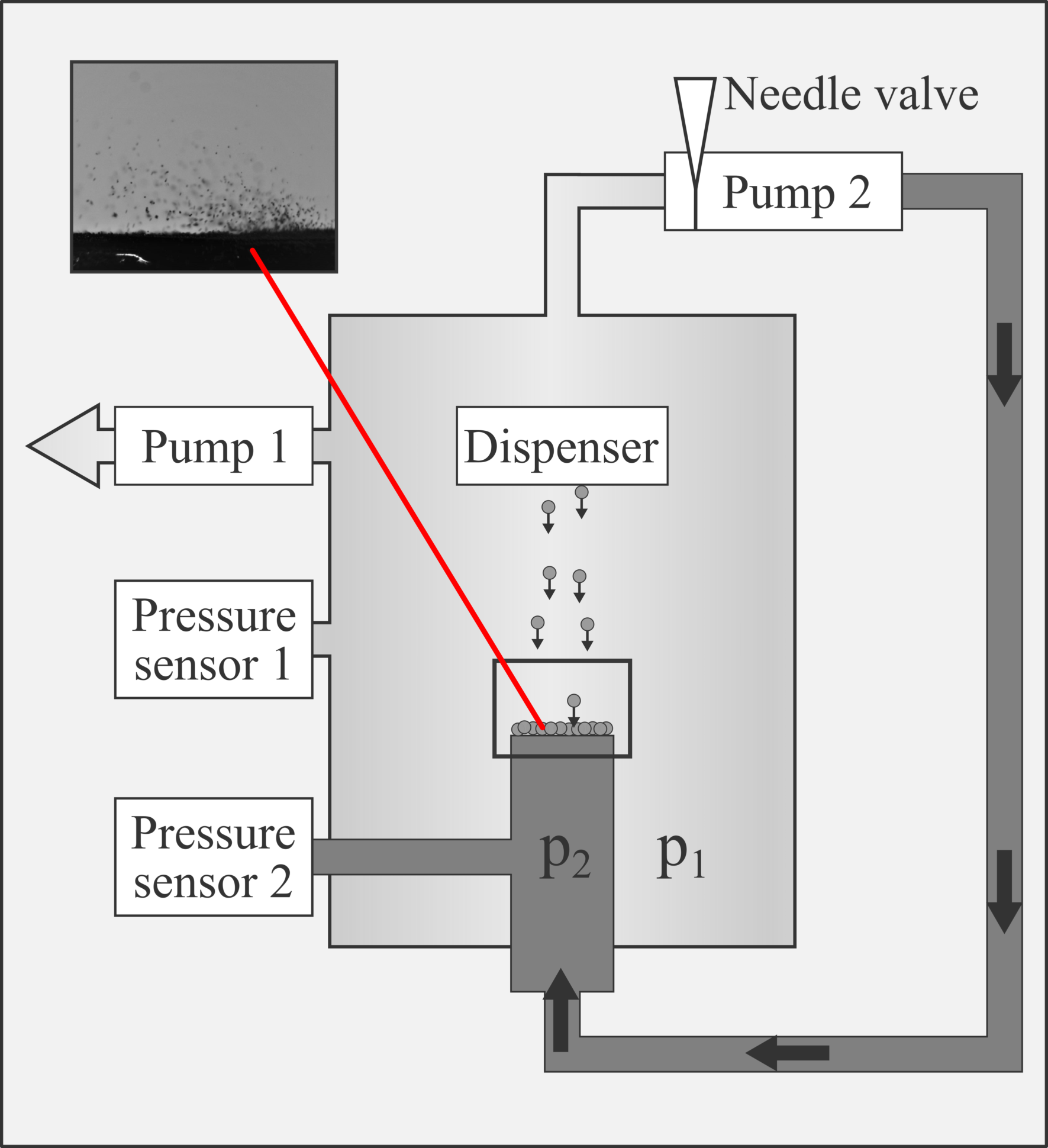}
\end{center}
    \caption{\label{fig.aufbau}Overall sketch of the experiment; A vacuum pump (left, pump 1) is used to set an ambient pressure of 600 Pa. Then particles are dispersed and settle on a mesh. Upon pumping gas through this mesh with a second pump (top right, pump 2) a pressure drop is set ($\rm P_2 > P_1$), measured by two pressure sensors. Particle motion is observed and recorded with a camera (example image as inset in the top left).}
\end{figure}

A sketch of the setup is shown in fig. \ref{fig.aufbau}. The setup is a vacuum chamber system, which is  evacuated to 600 Pa. In principle, the vacuum system is only a closed loop with a second pump within the loop. This pump can circulate gas through the system. The gas flow is controlled by a needle valve. The mesh mentioned above, which eventually holds the sample, is also part of the loop and any gas pumped has to pass through the mesh and sample. Therefore, once the circulating membrane pump is started, a pressure difference on both sides of the mesh and sample up to several Pa is established. The vacuum parts below and above the mesh are large volume vacuum chambers to avoid pressure drops due to small tubing. The pressure drop along the sample can be measured with two pressure sensors at these two vacuum chambers under and above the mesh and sample.

Nevertheless, a first procedure is measuring the pressure drop without particle sample. This quantifies the resistance of the system due to the flow in general. 
In detail the pressure drops are always taken as the difference between the two pressure sensors. Before the ciruclation is started, without gas flow, the offset between both sensors can be measured as there is no pressure difference initially. This way, the pressure drop along the mesh and the mesh plus sample can be measured to an accuracy of better than 0.6 and 0.8 Pa, respectively. The error originates in the pressure measurements. We used two CMR 362 sensors (Pfeiffer Vacuum) with a resolution of 0.3 Pa. {The pressure difference adjusts itself on a timescale of seconds to about a minute and ejections occur during the total adjustment time depending on the current pressure difference. 
Each experimental cycle from evacuating the setup to pressurizing it again to remove the current particle bed takes about 1h.
}

The thickness of the particle layer at the local point of particle ejection occuring is determined from a differential image between particle holder without sample and with sample. The spatial resolution of the optical system is 10 $\rm \mu m$. {Between the measurement cycles venting a vacuum chamber is necessary to remove the dust layer from the previous measurement.}
\subsection{Particle samples}
\begin{figure}
\includegraphics[width=\columnwidth]{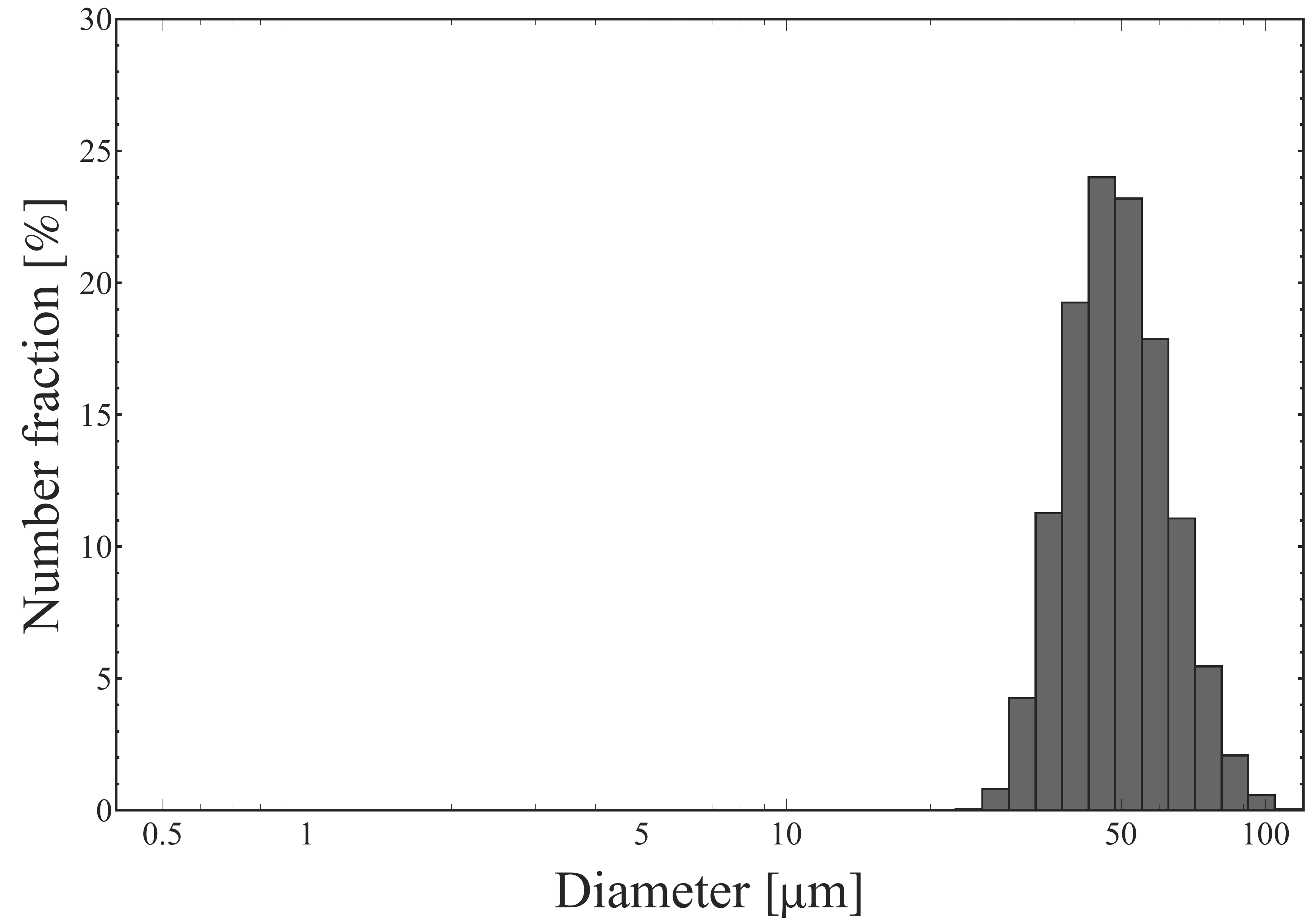}
\includegraphics[width=\columnwidth]{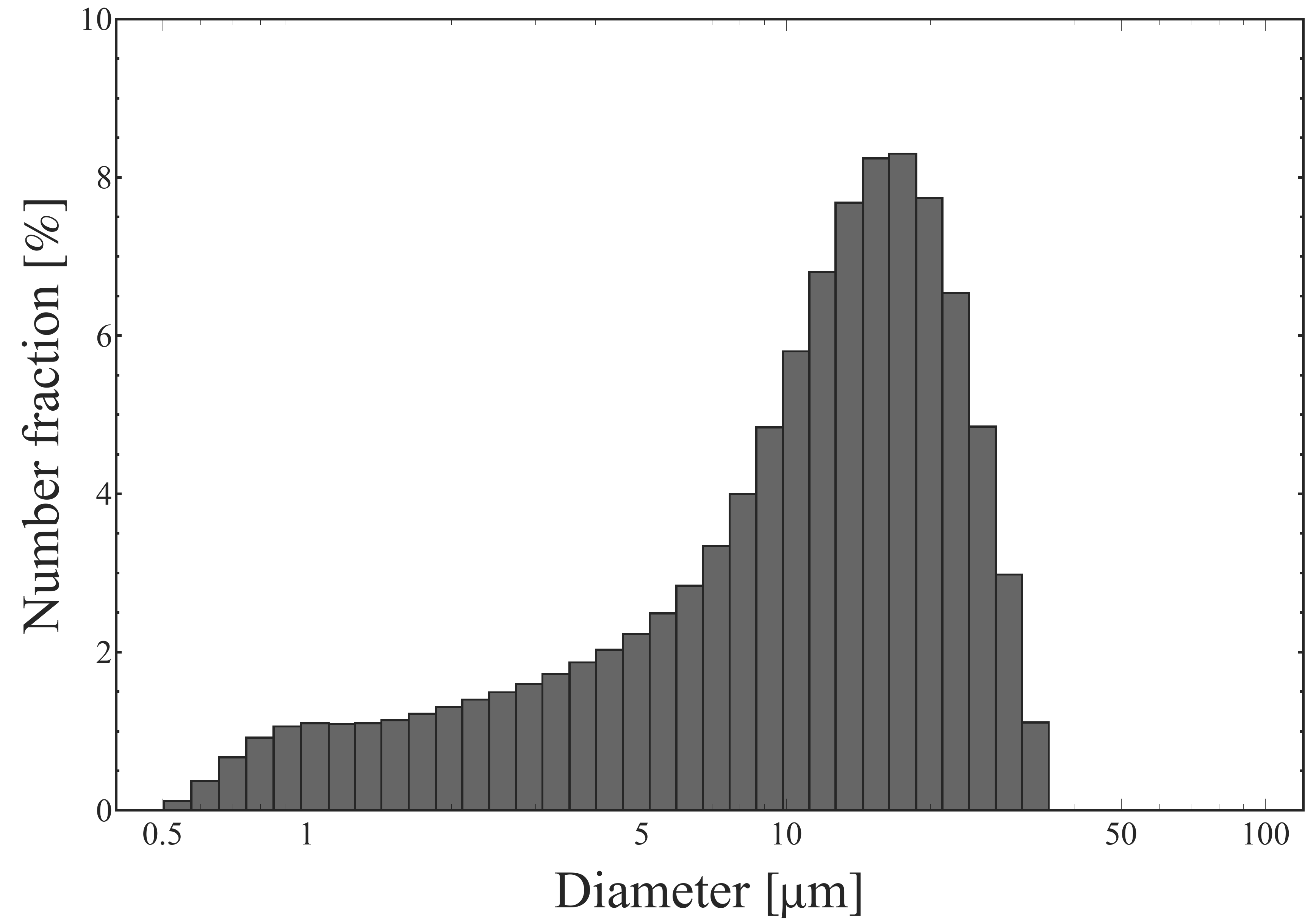}
    \caption{\label{fig.sizes} Size distribution of the glass beads (top) and basalt grains (bottom) as measured by a commercial Mastersizer 3000 (Malvern).}
\end{figure}

\begin{figure}
\includegraphics[width=\columnwidth]{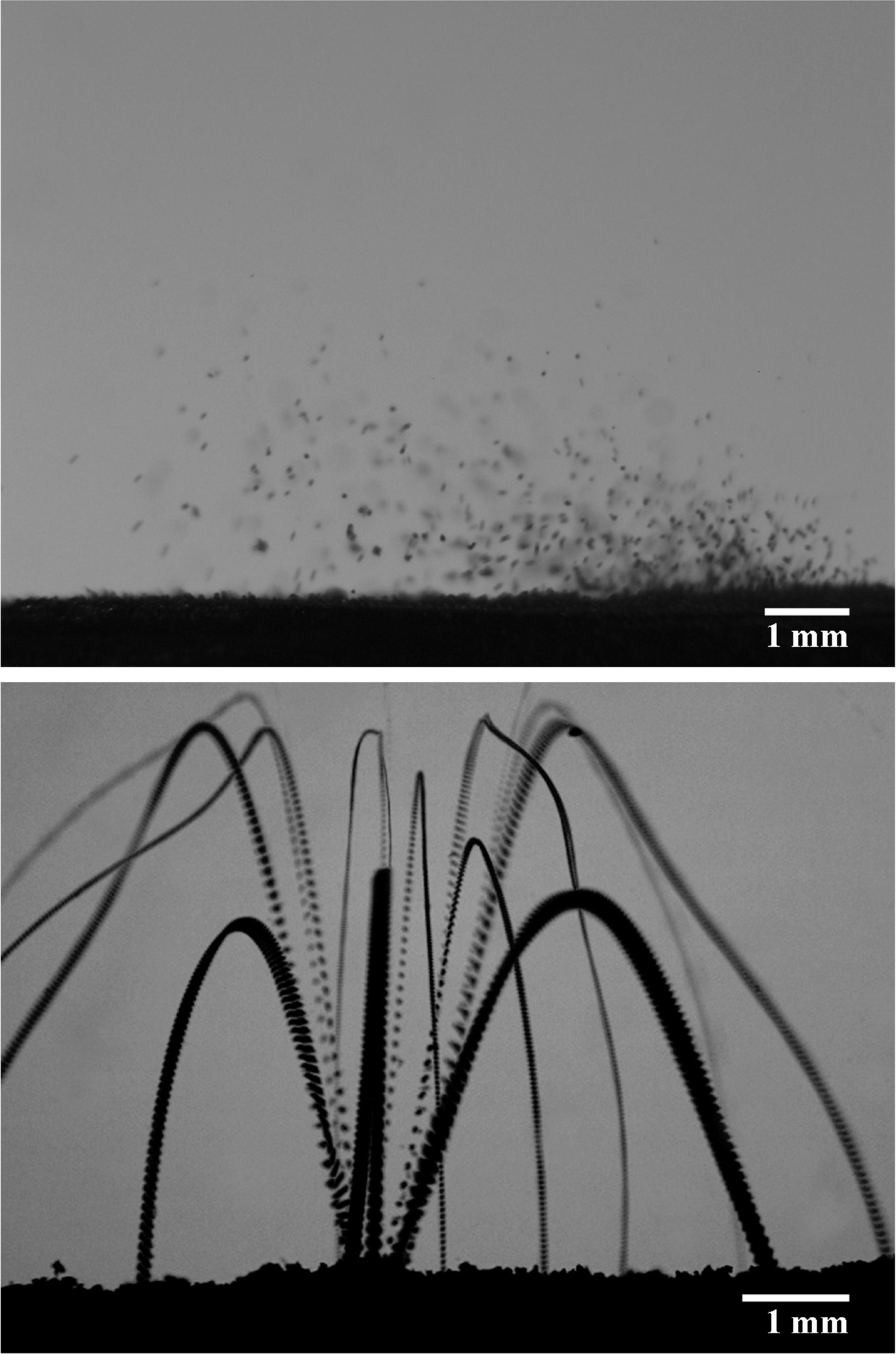}
    \caption{\label{fig.lifting} Two examples of lifting events; Top: Glass particles ejected during one event, single frame {from slow frame rate camera}; Bottom: Basalt particles ejected in a more localized event. {Here, we specifically focus on the trajectories and images were taken with a different camera at high frame rate of 1438 frames per second}. The images were superimposed on each other by considering the lowest value at each pixel (darkest pixels add up).  Images are contrast enhanced.}
\end{figure}

We used two different particle samples here. The first sample consists of spherical glass particles. 
The grain size peaks at about 50 $\rm \mu m$. This implies that the thinnest layers which we prepared ($\sim $50 $\rm \mu m$) are only a monolayer. The detailed size distribution of the grains is shown in fig. \ref{fig.sizes} (top).
The second sample consists of basalt grains. They are irregular and somewhat smaller with typical sizes of about 20 $ \rm \mu m$. Here, the smallest layer prepared (100 $\rm \mu m$) consists of at least a few particle layers. A detailed size distribution is given in fig. \ref{fig.sizes} (bottom) and also includes a significant amount of smaller grains of only a few micrometer, which might be closer to the conditions on the surface of Mars.
Two examples of observed grain lifting events, one for each particle sample, are shown in fig. \ref{fig.lifting}. {As example with the camera used at low frame rate to detect ejections,  a single frame of an ejection is shown for glass in fig. \ref{fig.lifting} top. High speed videos with a different camera were recorded to visualize the trajectories. A superposition of basalt images is shown in fig. \ref{fig.lifting} bottom. A superposition of images from a high speed video of the glass samples is also shown in fig. \ref{fig.saltating} below. }

\section{Results}

\begin{figure}
\includegraphics[width=\columnwidth]{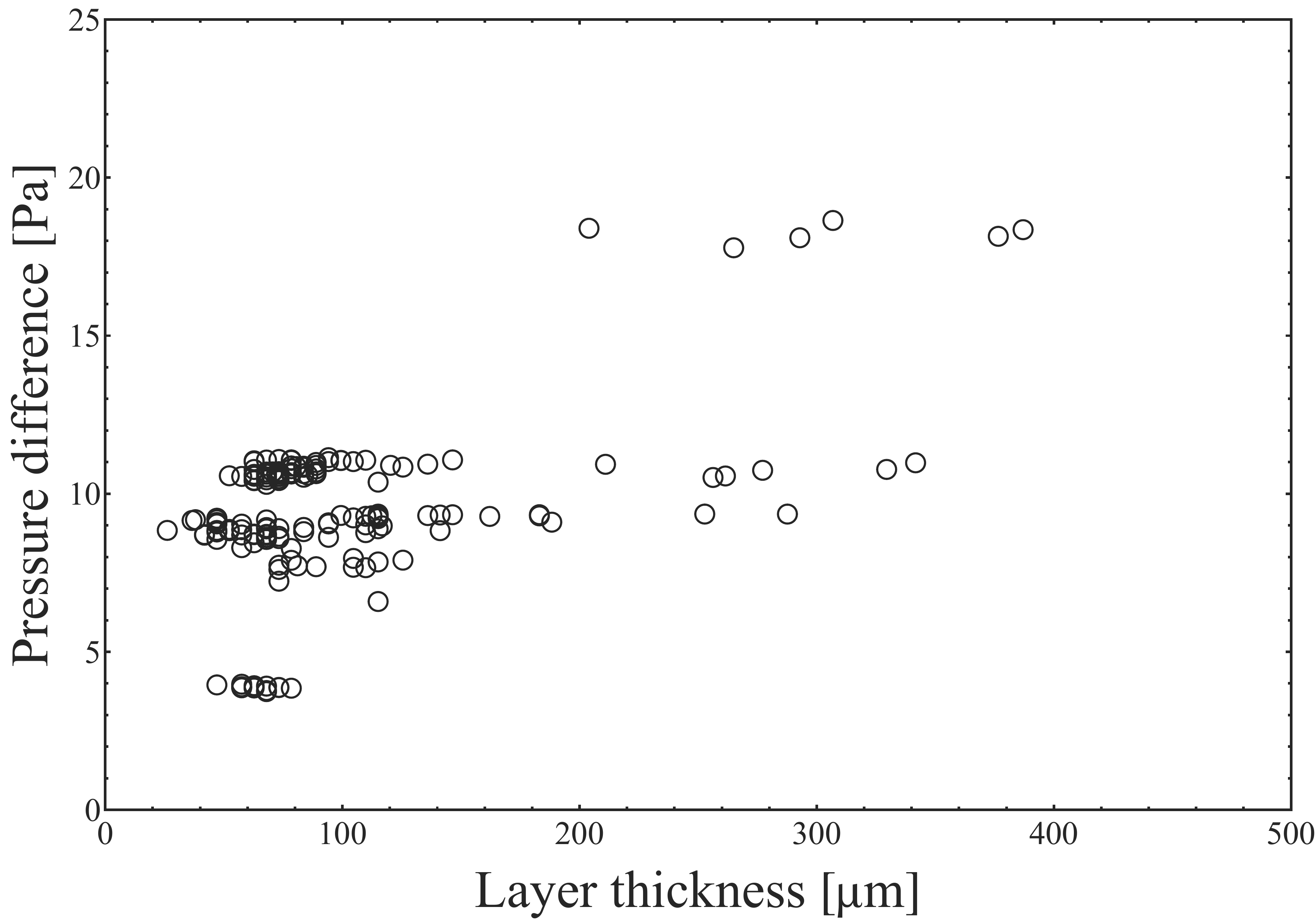}
\includegraphics[width=\columnwidth]{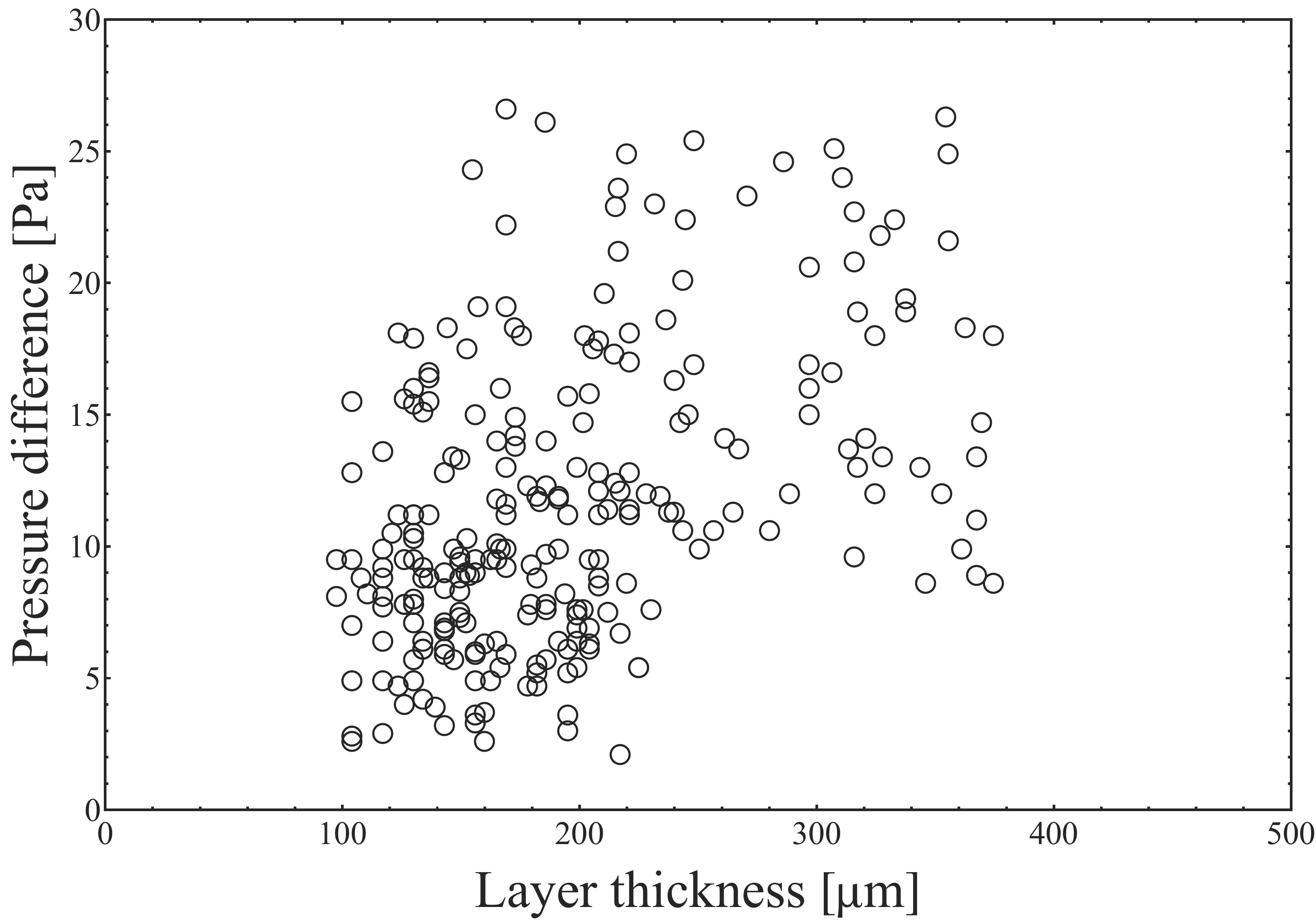}
    \caption{\label{fig.deltapg}Pressure differences needed to eject particles. Top: Glass beads; Bottom: Basalt grains}
\end{figure}

\begin{figure}
\includegraphics[width=\columnwidth]{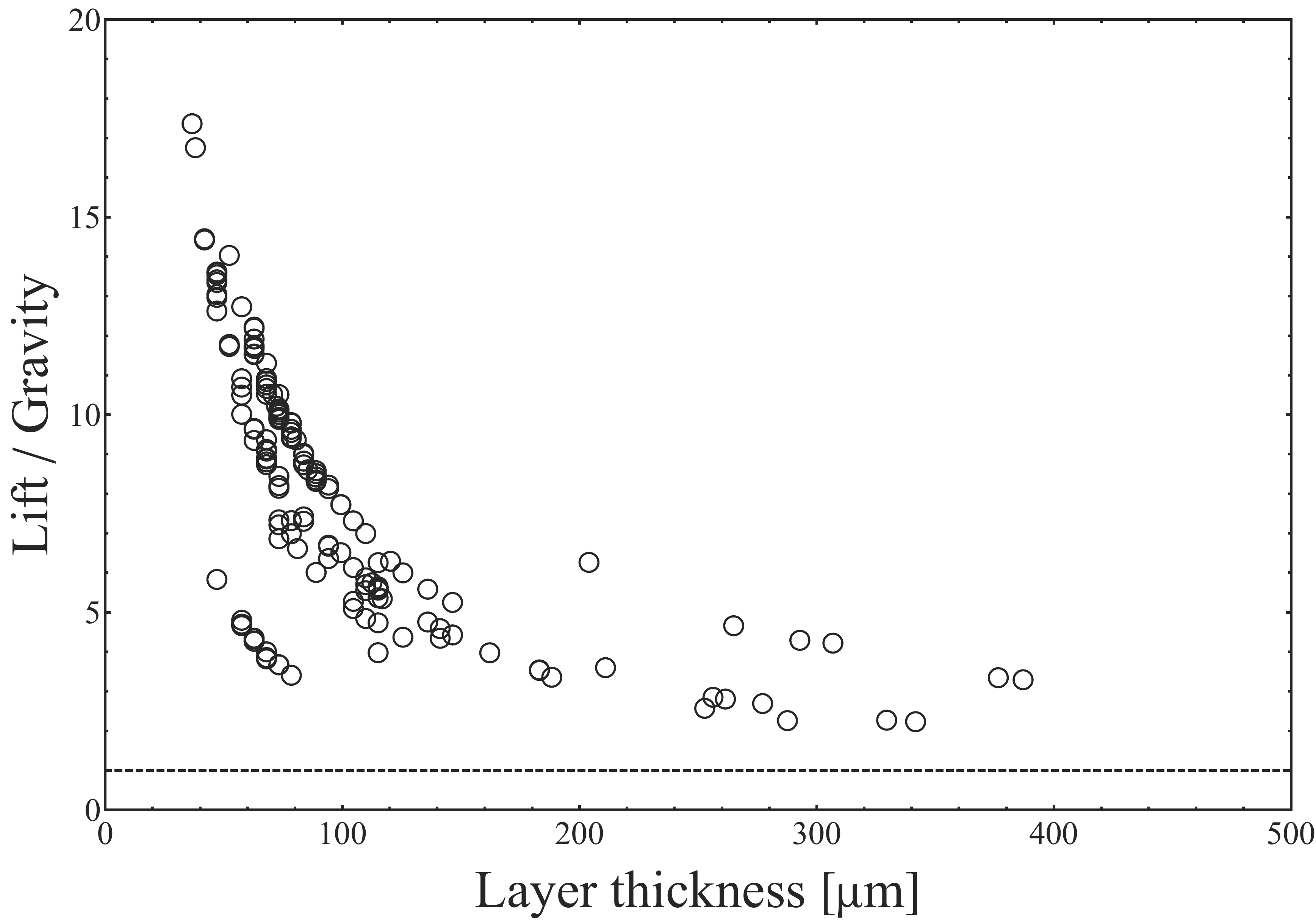}
\includegraphics[width=\columnwidth]{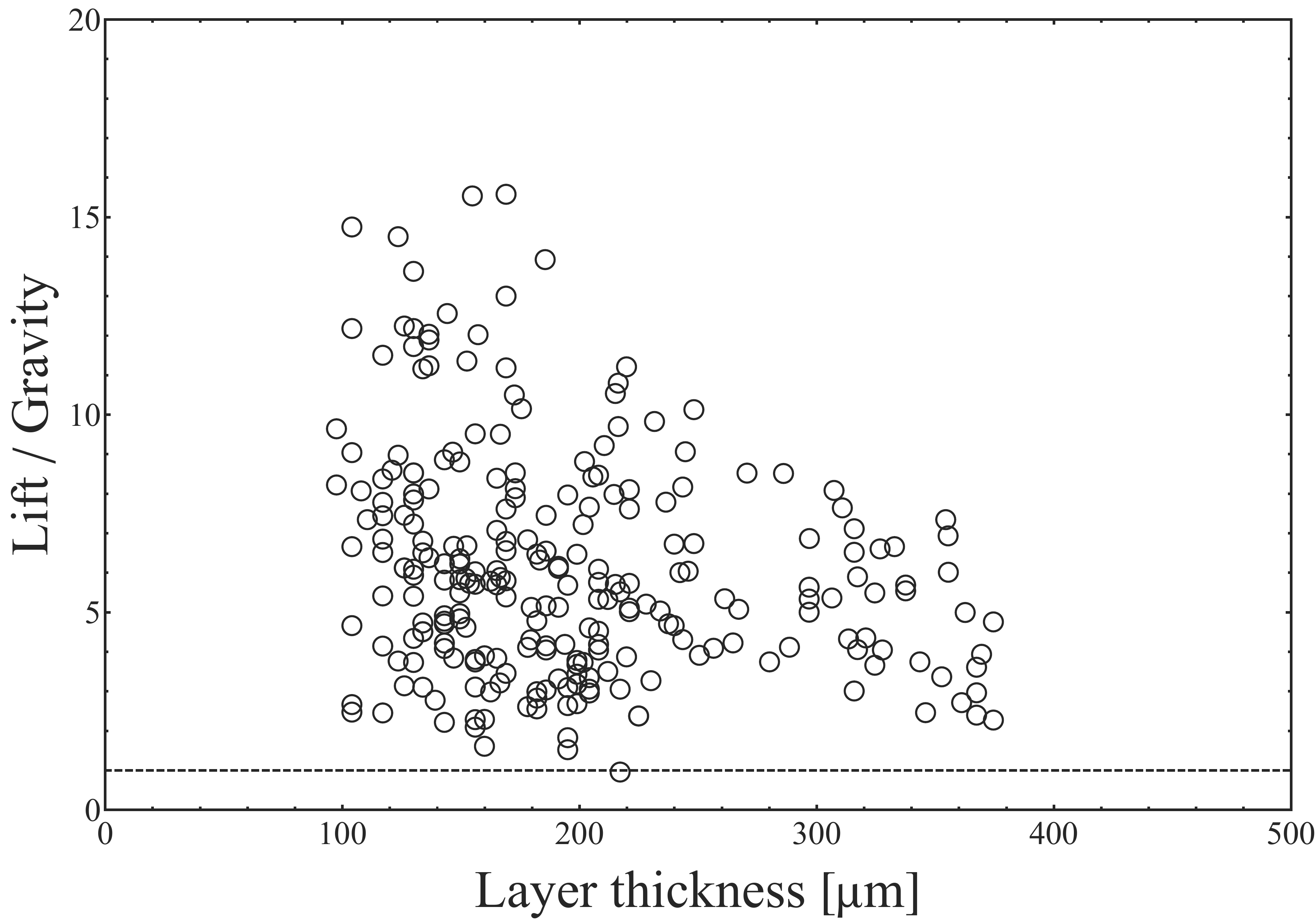}
    \caption{\label{fig.deltapgrav}Ratio between pressure difference at particle ejection and hydrostatic pressure difference due to the weight of the particle layer. Top: Glass; Bottom: Basalt; The dashed lines mark 1, where the pressure difference equals the hydrostatic pressure differnce due to the weight of the layer and cohesion does not contribute significantly.}
\end{figure}

The main result is the pressure difference needed to trigger the first particle ejections. This is shown in fig. \ref{fig.deltapg} depending on the layer thicknesses for glass beads as well as basalt grains. 
Up to 200 $\rm \mu m$ thickness for basalt, a few Pa are sufficient to lift grains with 2.0 $\pm 0.8$ Pa, being the lowest value measured. 
Fig. \ref{fig.deltapgrav} also gives the ratio between pressure difference and hydrostatic pressure difference due to the weight of the particle layer calculated as $\rho g d$, with layer mass density $\rho$, gravitational acceleration on Earth $g$, and layer thickness $d$. 
{We used a layer density of 1030 $\rm kg/m^3$ for basalt and 1468 $\rm kg/m^3$ for glass measured for large sample masses just poured into a recipient to accurately measure volume and mass.} Detailed values will depend on the volume filling factor, which might vary by some significant factor below 2. {The data in fig.  \ref{fig.deltapg} and \ref{fig.deltapgrav} originate from 9 experimental runs for glass beads and 5 experimental runs for basalt. Each experiment cycle provides a few tens of observed ejection events. Each data point symbolizes a single ejection within one of the measurement cycles.}

For small layers cohesion usually dominates over gravity by a factor of a few up to a few tens. As the thickness gets larger, it is more and more gravity that has to be compensated for, as also found by \citet{Koester2017a}. The role of cohesion decreases then. 

The glass beads are spherical, relatively large, almost sand size, and monodisperse. They likely arrange in regular patterns. This explains the "steps" for glass bead layers, where individual runs of the experiment essentially show constant cohesion.  
However, this is different for basalt due to the variable nature of cohesion for irregular, smaller grains with wider size distribution. At low thickness, some ejections only required gravity to be compensated. So cohesion is not dominating in these cases. 

For cohesive particles, the reduced gravity on Mars would not help lifting grains. However, the first grains ejected are not attached to the ground by cohesion. Only gravity has to be balanced and even pressure differences below 1 Pa might be sufficient at Martian gravity to lift grains.
 
It should be noted that our samples were prepared by sieving. On Earth the impacts during preparation are much more intense than on Mars. \citet{Musiolik2018} found that samples prepared under Martian gravity can have strongly reduced cohesion. So also here, it is likely that a few Pa as occuring in dust devils might actually be sufficient not only for lucky winners but for a significant amount of grains.

Lifting grains at all is also important in view of an additional result.
Ejected grains regularly return to the sample after they are free and inevitably lose their pressure support. This particle return has two aspects. 

First, it tells us that it is mostly static pressure support and not gas drag by the flow that initiates the particle motion. 
{This is evident as, in contrast to the pressure difference, gas drag would still be present once grains are released from the surface and in a homogeneous flow, grains just cannot return against this gas flow. However, as seen below some of these grain impact the surface again. That does not mean that the gas flow is not important for the further transport of some grains especially as the ejection separates grains of different sizes. Some details of this can be deduced from the trajectories  seen in fig. \ref{fig.lifting}, bottom. The small particles show trajectories with sideward motion, only sedimentating after some prolonged stay at a certain height. This clearly shows that these particles experience a gas drag on the order of gravity. Not shown here, very small fragments couple so well to the gas flow that they are even leaving for good with the gas flow.} 

{The large particles behave differently. In general, particles in a gas flow under gravity have a damped motion \citep{Wurm2001}. They follow any change on a gas-grain friction timescale $\tau$ and then move with a constant speed $v$ relative to the gas with $v=\tau g$, where $g$ is the gravitational acceleration. The large grains in our sample of $\sim 100 \rm \, \mu m$ (big particles in fig. \ref{fig.lifting} bottom) have gas-grain coupling times at 600 Pa on the order of $\tau \sim 10 \, \rm ms$ \citep{Blum1996}. With the given frame rate, particles are imaged more often than every ms and  particles should be accelerated for a number of frames. At least the spacing between the first few images on the superposition in fig. \ref{fig.lifting} bottom should increase. This is not the case. In contrast, particles have a very high initial velocity (large spacing between first frames) but they remain with that speed or even slow down slightly right away even if they are only about 100 $\rm \mu m$ above the surface. As we only consider the gas flow to be dispersed on the mm-scale due to the extension of the particle layer, this motion cannot be achieved by a simple gas flow. On the other side, a high initial velocity and immediate slow down afterwards is expected if a pressure release initially accelerates the particles. In summary, in a sense this is similar to saltation, where large grains are driven by the wind and small dust liberated in these events is prone to wind transport. }

\begin{figure}
\begin{center}
\includegraphics[width=\columnwidth]{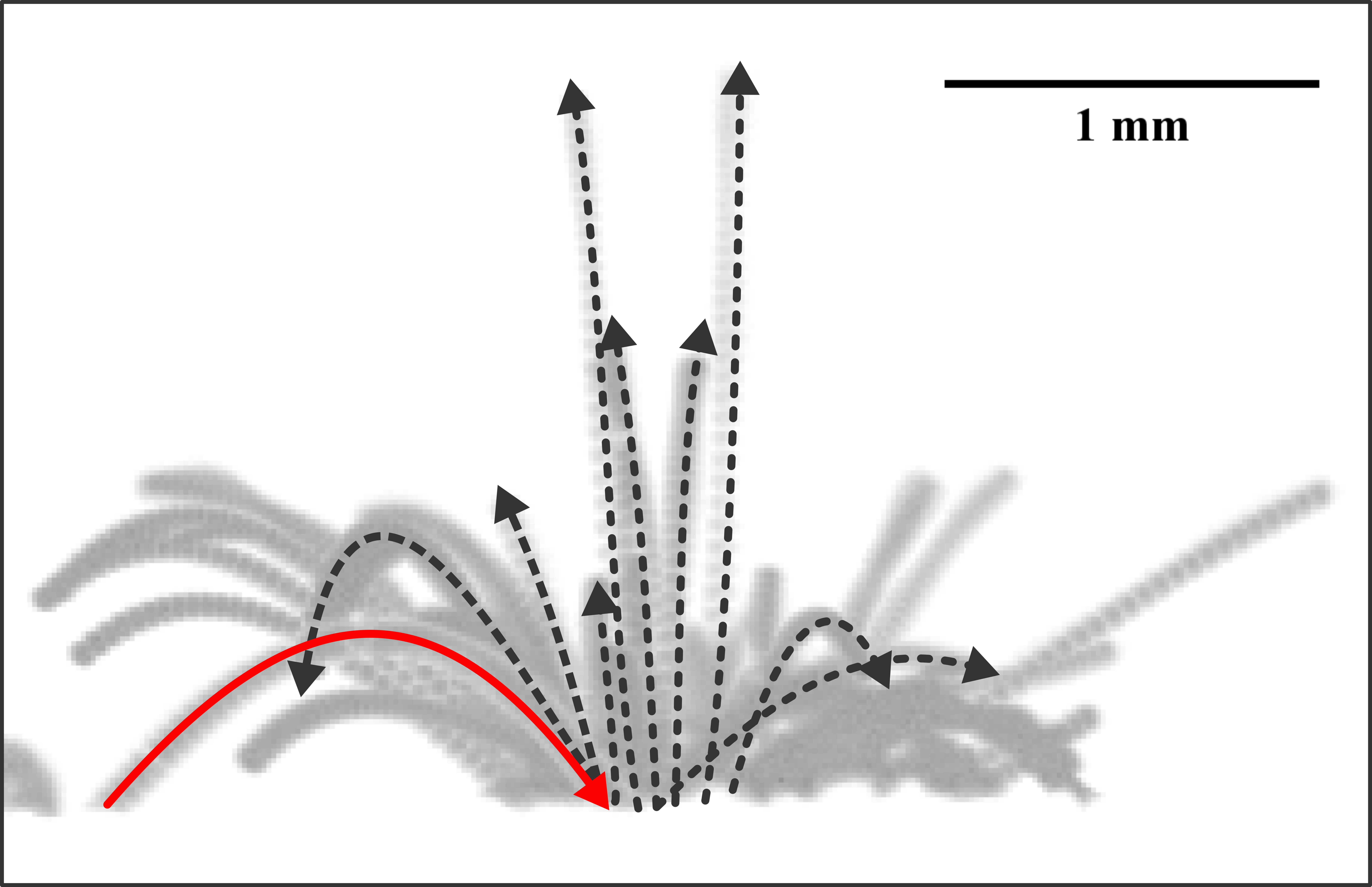}
    \caption{\label{fig.saltating} Example of impact triggered ejection of glass beads. Trajectories are superimposed images of a high speed video (1438 fps). Solid red arrow: A single particle ejected at the left returns to the particle bed. Dashed black arrows: Tracing some examples of ejecta formed after impact of the single grain.}
\end{center}
\end{figure}

As second aspect, the grains which collide again with the particle bed sometimes kicked off new grains as shown in fig. \ref{fig.saltating}. If we borrow the term \textit{saltation} again, this looks similar to saltation but has a different mechanism. The re-impact velocity would be much too low and the particles cannot be lifted by momentum transfer alone \citep{Bogdan2019}. This is also evident as the  energy of the single grain is only a very small fraction of all particle energies of ejecta produced. Here, as avalanche mechanism, only some cohesion has to be lowered by the impact. The pressure difference then lifts the grains. This shows that the total dust layer is under strong tension close to be compensating gravity and cohesion and only needs a small trigger to erupt. In fact, the example shown in fig. \ref{fig.lifting} top is just one frame from an image sequence which shows a large wave of particle ejections moving to the right. 

\section{Conclusion}

For the first time, we could apply pressure differences of only a few Pa to only 100 $\mathrm{\mu m}$ thin layers of dust (and sand) sized grains to study grain lifting. We found that first grains start to lift at only 2.0 $\pm 0.8$ Pa and more grains typically are lifted at about 10 Pa. The lower end is in the range of pressure drops observed in Martian dust devils {according to the reported values of 0.3 to 5 Pa \citep{Murphy2002, Ellehoj2010, Kahanpaeae2016, Steakley2016}. The higher values around 10Pa might be met on Mars as well as indicated by first results of the InSight mission \citep{Banerdt2019}.}

It is mandatory for the lift that there is a large enough gas reservoir below the layer and / or low enough flow rate through the layer to sustain the pressure difference. This concern on degassing was already phrased by \citet{Balme2006}. It did not vanish and we cannot resolve that concern here. If the grains are only part of a thicker but homogeneous layer and even if this layer is only a mm thick, pressure differences would not be strong enough to lift grains but that conclusion was obvious from the start. 

Preparing artificial layers on a stable mesh, our laboratory experiments are still far from reality.
We could have disproven the mechanism if we had found no ejections at all in the relevant pressure difference range. This did not happen. Therefore, while cohesion gains importance the smaller the layers get, it is not a limiting factor for particle lift.


To conclude, the experiments did not yet prove, ultimately, that the $\Delta P$-effect works under real Martian conditions. The evidence is strong though. {Under} ideal conditions, this effect alone can lift grains without support from any other lifting mechanisms and without wind. With that in mind, it might still contribute significantly to lift if conditions are not as ideal. Therefore, we conclude that the results clearly show the potential of the $\Delta P$-effect as a powerful lift agent in Martian dust devils. 

\section{Acknowledgements}
We thank Fr\'{e}d\'{e}ric Schmidt and the anonymous referee for 
a constructive review.

\bibliography{bib}

\end{document}